%% file: paper.tex
\documentclass[10pt,sigconf,letterpaper,nonacm]{acmart}
\pdfoutput=1
\usepackage{amsmath}
\usepackage{amsthm}
\usepackage{amsfonts}
\usepackage{graphicx}
\usepackage{pifont}
\usepackage[font={small,it}]{caption}
\usepackage{subcaption}
\usepackage{seqsplit}
\usepackage{tabularx}

\usepackage{color}    
\usepackage{url}
\usepackage{mathrsfs}
\usepackage{msc}
\usepackage{pgf-pie}
\usepackage{threeparttable}
\usepackage{booktabs} 
\usepackage{multirow}
\usepackage{tabularx} 
\usepackage{array, makecell}
\newcolumntype{L}[1]{>{\raggedright\arraybackslash}p{#1}}
\newcolumntype{C}[1]{>{\centering\arraybackslash}p{#1}}
\newcolumntype{R}[1]{>{\raggedleft\arraybackslash}p{#1}}
\usepackage{pgfplots}
\usepackage{scalefnt}
\usepackage[linesnumbered,commentsnumbered,ruled,vlined]{algorithm2e}
\usepackage{paralist}
\usepackage{tikz}
\pgfdeclarelayer{bg}
\pgfsetlayers{bg,main}
\usetikzlibrary{decorations.text} 
\usetikzlibrary[arrows,decorations.pathmorphing,backgrounds,positioning,fit,matrix]
\usetikzlibrary{decorations.pathreplacing}
\usepackage{listings}

\usepackage{pifont}
\newcommand{\point}[1]{\par\smallskip\noindent\textbf{#1.}}

\makeatletter\def\@captype{table}\makeatother

\input{code/solidity-highlighting.tex}
\colorlet{punct}{red!60!black}
\definecolor{delim}{RGB}{20,105,176}
\colorlet{numb}{magenta!60!black}
\lstdefinelanguage{json}{
    numbers=left,
    numberstyle=\scriptsize,
    stepnumber=1,
    numbersep=8pt,
    showstringspaces=false,
    breaklines=true,
    literate=
     *{0}{{{\color{black}0}}}{1}
      {1}{{{\color{black}1}}}{1}
      {2}{{{\color{black}2}}}{1}
      {3}{{{\color{black}3}}}{1}
      {4}{{{\color{black}4}}}{1}
      {5}{{{\color{black}5}}}{1}
      {6}{{{\color{black}6}}}{1}
      {7}{{{\color{black}7}}}{1}
      {8}{{{\color{black}8}}}{1}
      {9}{{{\color{black}9}}}{1}
      {:}{{{\color{punct}{:}}}}{1}
      {,}{{{\color{punct}{,}}}}{1}
      {\{}{{{\color{delim}{\{}}}}{1}
      {\}}{{{\color{delim}{\}}}}}{1}
      {[}{{{\color{delim}{[}}}}{1}
      {]}{{{\color{delim}{]}}}}{1},
}
\AtBeginDocument{%
  \providecommand\BibTeX{{%
    \normalfont B\kern-0.5em{\scshape i\kern-0.25em b}\kern-0.8em\TeX}}}
\makeatletter
\def\@copyrightspace{\relax}
\makeatother

\begin{document}
\newcommand{\synt}{Synthetix\xspace}%
\newcommand{\maker}{MakerDAO\xspace}%
\newcommand{\cpd}{Compound\xspace}%
\newcommand{\af}{AmpleForth\xspace}%
\newcommand{\cl}{Chainlink\xspace}%
\fancyhead{}

\title{A First Look into DeFi Oracles}
\author{Bowen Liu}
\affiliation{%
  \institution{SUTD, Singapore}
}
\email{bowen\_liu@mymail.sutd.edu.sg}

\author{Pawel Szalachowski\#}
\affiliation{%
  \institution{SUTD, Singapore}
  \country{\#Now at Google.}
}
\email{pjszal@gmail.com}

\author{Jianying Zhou}
\affiliation{%
  \institution{SUTD, Singapore}
}
\email{jianying\_zhou@sutd.edu.sg}

\begin{abstract}
    Recently emerging Decentralized Finance (DeFi) takes the promise of
    cryptocurrencies a step further, leveraging their decentralized networks to
    transform traditional financial products into trustless and transparent
    protocols that run without intermediaries.  However, these protocols often
    require critical external information, like currency or commodity exchange
    rates, and in this respect they rely on special \textit{oracle} nodes.  
    In this paper, we present a comprehensive measurement study of DeFi price 
    oracles deployed in practice.  
    First, we investigate designs of mainstream DeFi platforms that
    rely on data from oracles. We find that these designs, surprisingly, position
    oracles as trusted parties with no or low accountability.  Then, we present
    results of large-scale measurements of deployed oracles. We find and report that
    prices reported by oracles regularly deviate from current exchange rates,
    oracles are not free from operational issues, and their reports include
    anomalies.
    Finally, we compare the oracle designs and propose potential improvements.
\end{abstract}



\maketitle

\section{Introduction}
\label{sec:intro}
\input{sec/intro}

\section{Background}
\label{sec:background}
\input{sec/background}

\section{Measurements}
\label{sec:measurement}
\input{sec/measurement}

\section{Discussion}
\label{sec:discussion}
\input{sec/discussion}

\section{Related work}
\label{sec:related}
\input{sec/related}

\section{Conclusions}
\label{sec:conclusions}
\input{sec/conclusions}

\section*{Acknowledgment}
We gratefully acknowledge the anonymous reviewers for their valuable comments and suggestions,
and appreciate Michael Yiming Tan for helpful technical discussions about data collection.
This research is supported by the Ministry of Education, Singapore, under its
MOE AcRF Tier 2 grant (MOE2018-T2-1-111). 

\bibliographystyle{ACM-Reference-Format}
\bibliography{paper}

\end{document}

%% file: code/solidity-highlighting.tex
\usepackage{listings, xcolor}

\definecolor{verylightgray}{rgb}{.97,.97,.97}

\lstdefinelanguage{Solidity}{
	keywords=[1]{anonymous, assembly, assert, balance, break, call, callcode, case, catch, class, constant, continue, constructor, contract, debugger, default, delegatecall, delete, do, else, emit, event, experimental, export, external, false, finally, for, function, gas, if, implements, import, in, indexed, instanceof, interface, internal, is, length, library, log0, log1, log2, log3, log4, memory, modifier, new, payable, pragma, private, protected, public, pure, push, require, return, returns, revert, selfdestruct, send, solidity, storage, struct, suicide, super, switch, then, this, throw, transfer, true, try, typeof, using, value, view, while, with, addmod, ecrecover, keccak256, mulmod, ripemd160, sha256, sha3}, 
	keywordstyle=[1]\color{blue}\bfseries,
	keywords=[2]{address, bool, byte, bytes, bytes1, bytes2, bytes3, bytes4, bytes5, bytes6, bytes7, bytes8, bytes9, bytes10, bytes11, bytes12, bytes13, bytes14, bytes15, bytes16, bytes17, bytes18, bytes19, bytes20, bytes21, bytes22, bytes23, bytes24, bytes25, bytes26, bytes27, bytes28, bytes29, bytes30, bytes31, bytes32, enum, int, int8, int16, int24, int32, int40, int48, int56, int64, int72, int80, int88, int96, int104, int112, int120, int128, int136, int144, int152, int160, int168, int176, int184, int192, int200, int208, int216, int224, int232, int240, int248, int256, mapping, string, uint, uint8, uint16, uint24, uint32, uint40, uint48, uint56, uint64, uint72, uint80, uint88, uint96, uint104, uint112, uint120, uint128, uint136, uint144, uint152, uint160, uint168, uint176, uint184, uint192, uint200, uint208, uint216, uint224, uint232, uint240, uint248, uint256, var, void, ether, finney, szabo, wei, days, hours, minutes, seconds, weeks, years},	
	keywordstyle=[2]\color{teal}\bfseries,
	keywords=[3]{block, blockhash, coinbase, difficulty, gaslimit, number, timestamp, msg, data, gas, sender, sig, value, now, tx, gasprice, origin},	
	keywordstyle=[3]\color{violet}\bfseries,
	identifierstyle=\color{black},
	sensitive=false,
	comment=[l]{//},
	morecomment=[s]{/*}{*/},
	commentstyle=\color{gray}\ttfamily,
	stringstyle=\color{red}\ttfamily,
	morestring=[b]',
	morestring=[b]"
}

%% file: sec/intro.tex
One promise of open cryptocurrencies is to make money and payments universally
accessible without needing trusted parties.  Decentralized Finance (DeFi) aims
at extending this promise, proposing novel and traditional financial tools built
on the top of a blockchain-based smart contract platform.  DeFi offers multiple
advantages over traditional finance.  
First, it inherits the blockchain
properties, like decentralization, openness, accessibility, and
censorship-resistance. 
Second, DeFi is highly flexible, allowing for rapid
innovation and experiments by combining, stacking, or connecting different
financial instruments. 
Finally, DeFi provides interoperable services. 
Generally, new DeFi projects can be built or composed by combining other DeFi platforms.

An increasing trend within the DeFi ecosystem is hybrid protocols which try to
offer all advantages of DeFi, but eliminating the high volatility of 
cryptoassets~\cite{volatile},
which hinders broader DeFi adoption. They do so by connecting their cryptoassets 
to conventional financial instruments.
A prominent example is decentralized lending protocols that
have achieved more recent attention than any other category of DeFi.
\maker~\cite{maker}, a collateral-backed \textit{stablecoin} whose value is stable relative to USD, 
enables anyone to leverage their collateral assets to generate new tokens~\cite{SAI}
through a dynamic system of collateralized debts.
Once new assets are generated, they can be used in the same manner as any other
cryptocurrency. After paying down the debt and stability fee, the users
can withdraw collateral and close their loans.
Following the \maker success, other DeFi lending platforms,
like \cpd~\cite{compound}, were launched.
By offloading the traditional credit checking and reducing the costs with automation,
\cpd markets are actually a pool of assets that algorithmically derive interest-rates based on the supply and demand for the particular asset. Lenders and borrowers of these assets interact with the protocol directly in order to earn and pay respectively a floating interest rate, without having to negotiate any kind of terms such as maturity or interest rate.
As of March 2020, DeFi Pulse reports that
active outstanding loans from four open lending protocols — \maker,
Fulcrum~\cite{fulcrum}, dYdX~\cite{dydy}, and \cpd are above \$200 million~\cite{defipulse}.

Another example of a project that aims at value stability is \af~\cite{ampleforth}, the first DeFi protocol with supply elasticity. In response to the changes in demand, the platform always 
seeks a price-supply equilibrium based on the states of market and CPI index by
universally expanding to, or contracting from holders, aiming at making it robust to economic shocks and runaway
deflations.
\synt~\cite{synthetix}, another recent DeFi project, allows the creation
of ``synthetic assets'' -- \textit{Synths} whose prices can track currencies,
cryptocurrencies, and commodities.
The holders firstly lock their \synt native SNX tokens as collateral to
mint the Synths which are tokens intended to track the value of the target
asset (e.g., USD or gold).

All these systems require real-time information about the market price of the
assets used as collateral and redemption.  
It is necessary for their security as the value
(expressed in fiat) of cryptoasset collaterals is volatile.  
To implement this functionality, 
DeFi protocols introduce oracles, third parties reporting the
price of assets from real-world (off-chain) sources.
An oracle acts as a source of data that is being fed to a smart contract. 
Although oracles play a critical role in the
DeFi ecosystem, the underlying mechanics of oracles are vague and unexplored.
Firstly, their deployment practices, including how frequent the price updates,
how to aggregate the price value from multiple nodes, etc., are not transparent
nor accountable, leaving room for various misbehaving.  Secondly, the level of trust
placed in oracles is unclear and most likely unknown to many participants of the
ecosystem.  Finally, the impact of a potentially malicious oracle (or a group of
oracles) on the DeFi ecosystem is not investigated. 

In this work, we shed light on these issues, presenting the first comprehensive 
study on DeFi oracles.  First, we explain oracle designs
deployed in practice.  Second, we systemically investigate the deployment of
oracles for four popular open DeFi platforms - \maker, \cpd,
\af and \synt - which rely on external
oracles for price feeds.   We conduct detailed measurements on the price
deviation that comes from the differences between the information provided by
external oracles and real-world prices.  Moreover, we measure the robustness and
deployment practices of oracles by transaction graph analysis. Finally, 
we compare the deployed platforms, and we give insights on potential improvements.

%% file: sec/background.tex
Many DeFi protocols aim at providing low volatility of their cryptoassets by using
crypto collateral with its price pegged to some real-world assets.\footnote{We note,
that there are other designs that do not require pegs or
collaterals but these systems are out of scope this submission and we refer the
reader to recent surveys~\cite{moin2019classification,clark2019sok}.} 
Unlike in the real world, in DeFi protocols communicating exchange prices is not
trivial, since these protocols are implemented as smart contracts
deployed on the blockchain, without having access to any external resources
(like current asset prices). Therefore, in such
a design, price oracle is a fundamental component bridging cryptoassets with
the external information about their intended value.
In this section, we give a background on prominent DeFi protocols and their
oracle designs. 
All of these platforms, as well as the vast majority of all
DeFi platforms, are built upon Ethereum~\cite{ethdefiecosystem}.

\subsection{MakerDAO}
\label{sec:maker}
\maker is the most popular decentralized lending protocol where each its native
token DAI is pegged to USD and is backed by collateral in the form of
cryptoassets.  Since dealing with cryptocurrency volatility is a problem,
\maker offers the programmability of crypto without the downside of volatility
that you see with traditional cryptocurrencies like Bitcoin or Ethereum.  By
leveraging their cryptoassets into a Collateralized Debt Positions (CDP)
contract as collateral, users are able to generate multi-collateral DAI
\footnote{The single-collateral token in \maker earlier version is SAI.} tokens
that can be traded in the same manner as any other cryptocurrencies.  
In return,
the CDP accrues the debt known as overcollateralized loans determined
by collateralization ratio (i.e., C-Ratio), locking them out of access until
the outstanding debt is paid.
The C-Ratio currently is set to 150\% 
that helps the platform manage the risk of the borrowers by overcollateralizing the underlying assets.
When the users want to retrieve their collateral, they have to pay down the debt
in the CDP, plus the stability fee that continuously accrues on the debt over
time, which can only be paid in \maker's native tokens (MKR).  In addition to
payment of the stability fee, the MKR token also makes it possible to vote on
the evolution of the platform and plays an important role in the governance of
\maker, in proportion to the number of MKR each owner has.  The combination of
DAI as a stablecoin and MKR as a governance token ensures the equilibrium of the
system. Holders of MKR benefit directly from the use of DAI and the usage of the
DAI is managed by the holders who are able to protect the system.

\maker introduces an oracle module to obtain the real-time price of
assets. The accuracy of this information is critical as it determines whether a
CDP has enough collateral assets locked up and when a liquidation can be triggered.
\begin{figure}[h!]
	\centering
	\includegraphics[width=0.8\linewidth]{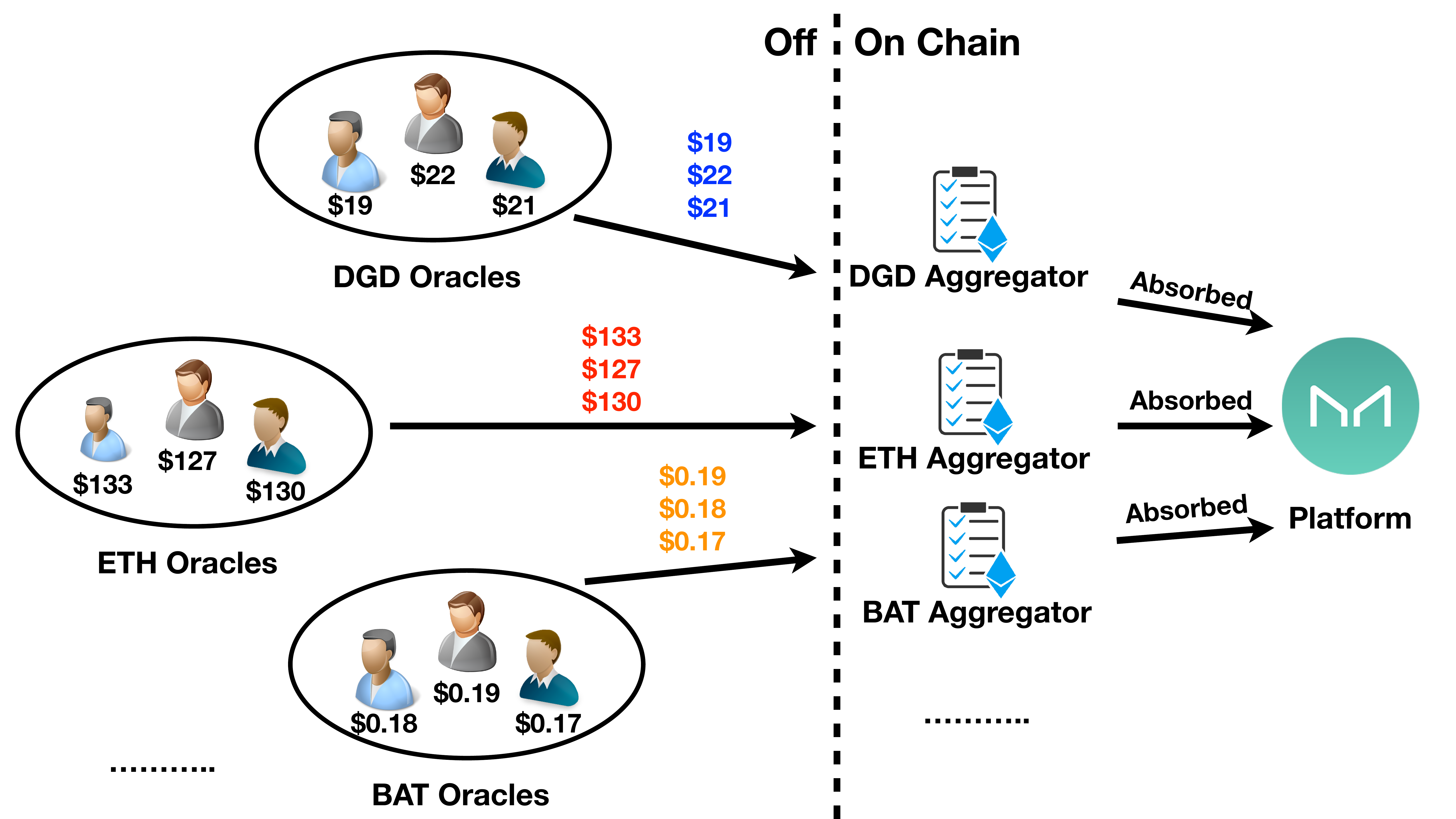}
    \caption{Oracle hierarchy of \maker platform.}
	\label{fig:makeroracle}
\end{figure}
The oracle module is composed of 
a number of whitelisted oracle addresses and an \textit{aggregator} contract.
Oracles send periodic price updates to the
aggregator which aggregates them, computes the median price as the
reference price, and
updates the platform by this reference price.
Each asset type has an independent aggregator contract
to collect information from authorized oracles.
We give a high-level overview of this architecture in Figure~\ref{fig:makeroracle}.
The aggregator contract implements access control logic allowing
addition and removal of price oracle addresses.
This operation is determined by the governors -- MKR holders, who
vote and update the changes on oracle addresses.
Moreover, the logic allows governors to set other parameters
that control the aggregator's behaviors, e.g., the minimum number of
oracles necessary to accept a new median value. 
Consequently, in such a decentralized
governance mechanism, the oracles could be manipulable
by MKR holders.  
Similar in style to 51\% attacks, 
a coalition
can profitably manipulate governance to ``steal'' 
system collateral~\cite{governanceattack}.

\begin{figure}[b!]
	\centering
	\includegraphics[width=0.7\linewidth]{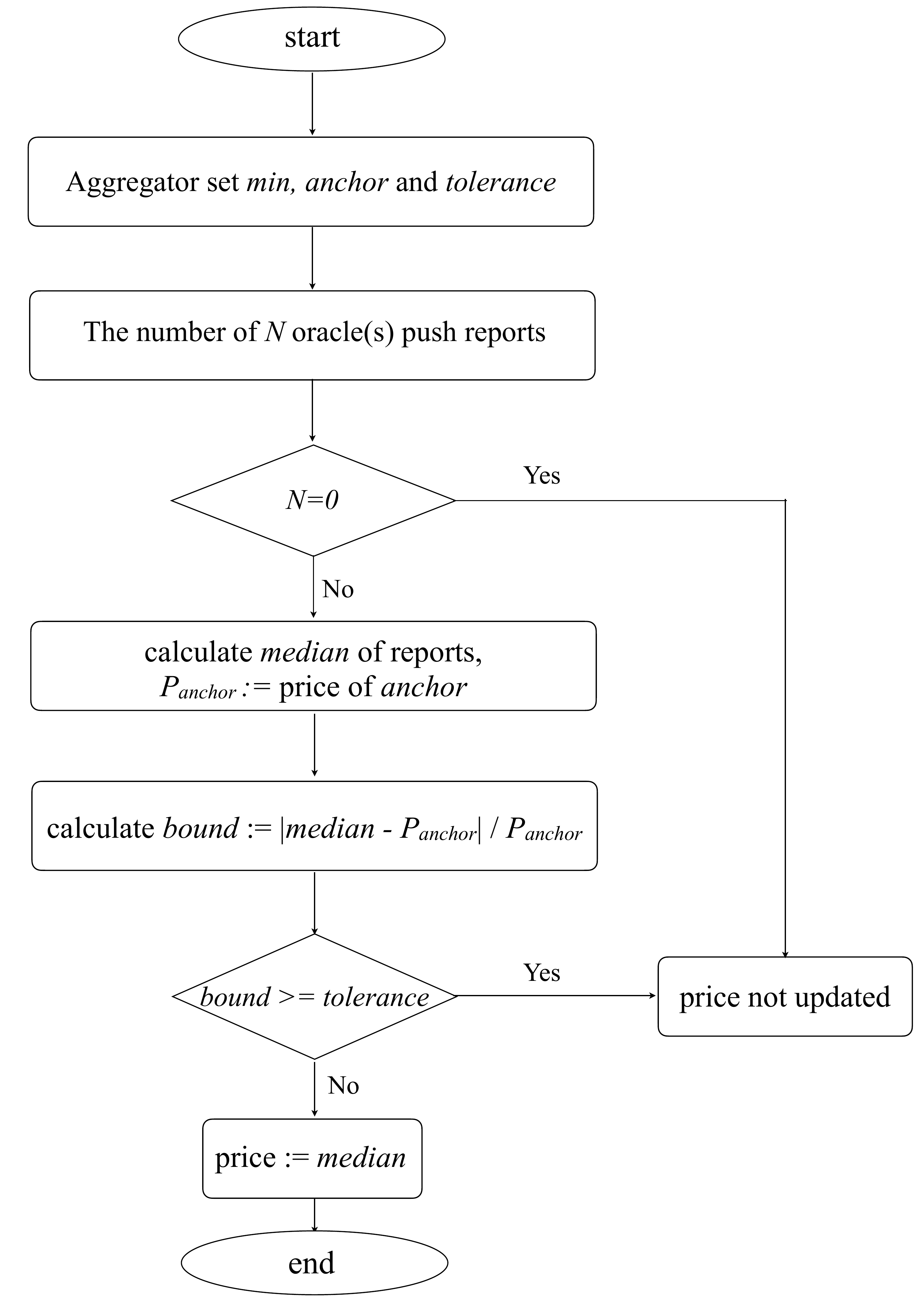}
    \caption{Oracle logics in \cpd.}
	\label{fig:oracle-interactions-compound}
\end{figure}
\subsection{Compound}
\label{sec:compound}
\cpd is a blockchain-based borrowing and lending platform where 
participants can lend their cryptoassets out and earn interest on it.  
Participants deposit their cryptoassets to the \cpd smart contract 
as collateral, and borrow against it. 
This contract automatically matches borrowers and lenders, and adjusts interest rates dynamically based on supply and demand.
Similar to \maker, \cpd employs oracles for price feed which are managed and controlled by
\textit{administrators} -- holders of \cpd's native COMP token.
\cpd platform is governed and upgraded by COMP holders
who propose, vote and implement any changes through the administrative functionalities. 
Proposals can include changes like adjusting an interest rate
model or collateral factor, managing the aggregator contract, and choosing 
the source of the oracle.

The logic of price updates in \cpd is depicted in
Figure~\ref{fig:oracle-interactions-compound} and as shown, at the beginning, the
administrator deploys an anchor contract and then creates an
aggregator contract with $min$, $anchor$ and $tolerance$ set, where
$min$ is the minimum number of reports necessary to calculate a new median value
which is set to one by default, $anchor$ indicates the address of anchor
contract and $tolerance$ rate is set to 10\%.  
The oracle
system in \cpd allows multiple authorized sources, known as reporters, to report
price data to the aggregator contract.  Reporters can be exchanges,
other DeFi projects, applications, over-the-counter (OTC) trading desks, etc.  
The
aggregator receives the reference prices from reporters, verifies them
and calculates the median value from them, and then stores it in order to be accessed by
the \cpd market.
The mechanism of updating the reference price of the asset is based on an
anchor price (reported by an anchor address), together with upper and
lower bounds that the median prices calculated by the aggregator are
checked against. Should the ratio between a new median price and the anchor
price be out-of-bounds, the official reference price of the asset would not be
updated.

\begin{figure}[b!]
	\centering
	\includegraphics[width=0.6\linewidth]{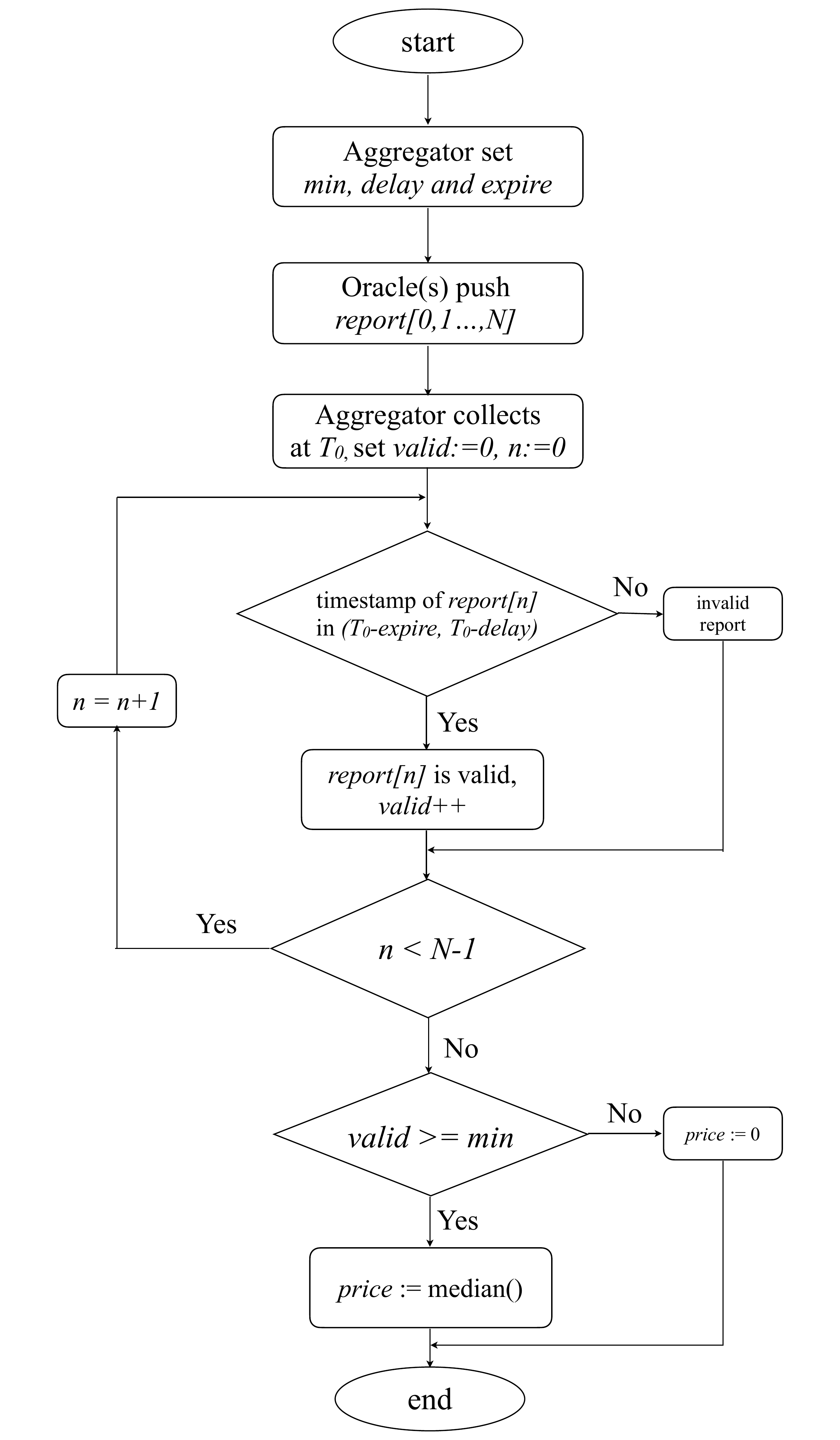}
    \caption{Oracle logics in \af.}
	\label{fig:oracle-interactions-ampleforth}
\end{figure}
\subsection{AmpleForth}
\label{sec:ampleForth}
Traditional commodities like gold or even cryptocurrencies like Bitcoin (produced based on a fixed supply schedule) cannot efficiently respond to changes in demand, making them vulnerable to destabilizing economic shocks and runaway deflation.
To address this shortcoming, \af creates AMPL tokens 
to automatically propagate asset price-information into supply.
By universally and proportionally expanding or contracting the quantity
of tokens from each holder based on the price exchange rate between the AMPL and USD, 
this automatic price-supply equilibrium is counter-cyclical and non-dilutive.
AMPL is initially pegged to USD, however not into perpetuity, since the
platform takes the Consumer Price Index into account to balance future USD inflation. 
Thus, \af aims for purchasing power stability by altering the supply based on the 
demand for AMPL tokens. 
More specifically, whenever there is more demand than supply, the platform automatically increases the total quantity of AMPL to prevent the relative price of goods from rising.
Similarly, when the total demand is less than supply, it decreases the quantity
accordingly.

It is critical for such protocol aiming at a price-supply equilibrium to have a reliable and accurate source of market
price information.
This core functionality of \af is depicted in
Figure~\ref{fig:oracle-interactions-ampleforth}.
The platform administrator sets
the $min$, $delay$, and $expire$ parameters of the aggregator contract during
its initialization, 
where $min$ (one by default), indicating the minimum number of providers 
with valid reports to consider the
aggregated report valid, 
$delay$ is the number of seconds since reporting that has to pass before a 
report is usable (set to one hour, by default),
and $expire$ represents the number of seconds after which the report is deemed expired.
In \af, this value is 12 hours by default.
A valid report must exist on-chain publicly for at least 1 hour before it can be used by the supply policy and will expire on-chain if a new report is not provided before 12 hours elapses.
That means that only the reports submitted within the \textit{valid range} are considered
as valid reports. 
\input{fig/valid-range-ample}
We depict this logic in Figure~\ref{fig:validreportrangeample} and let us assume that the aggregator retrieves price information at $T_0$.
The correct AMPL/USD price rate is median calculated by 
aggregator based on
the reports submitted by authorized oracles within the valid range.

\point{\cl}
Smart contract platforms, like Ethereum,
lack the ability of connecting smart contracts with off-chain resources
(like web) natively. 
\cl~\cite{chainlink} aims to resolve this issue by acting as a decentralized oracle network
bridging on-chain smart contracts with the off-chain environment.
(In Section~\ref{sec:related}, we discuss competing designs to \cl.)
It implements this by giving smart contracts APIs allowing to request 
off-chain resources such as market data, bank payments, retail payments,
back-end systems, events data, or web content.
\cl consists of a network of multiple decentralized, and independent oracles and
aggregators that collect and process off-chain data and deliver it (processed)
to smart contracts on request.
\af is an example of a platform that is integrated with \cl~\cite{chainlink}.

\subsection{Synthetix}
\label{sec:synthetix}
\begin{figure}[h!]
	\centering
	\includegraphics[width=0.9\linewidth]{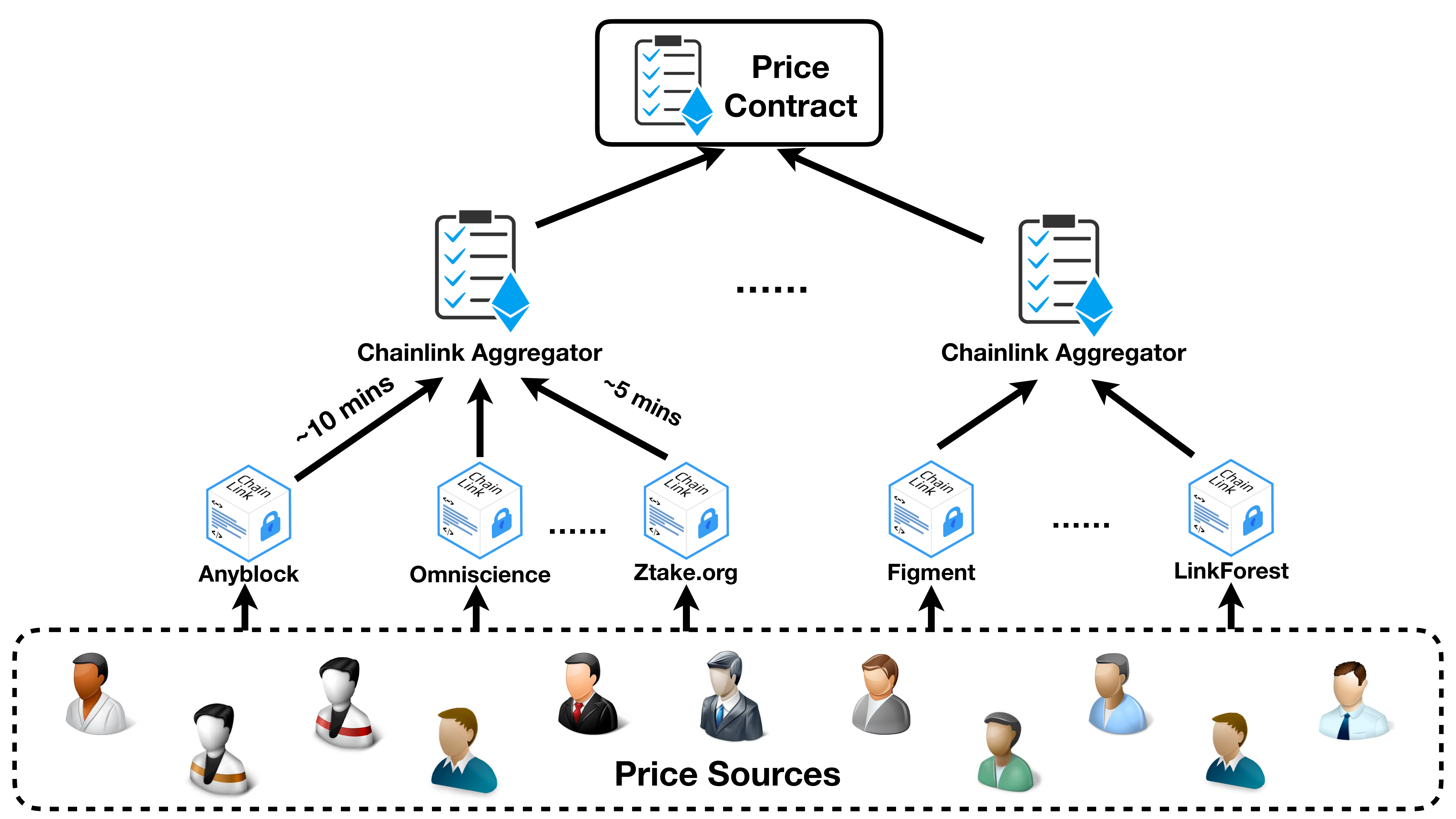}
    \caption{Oracle hierarchy of \synt platform.}
	\label{fig:synthetix}
\end{figure}
\synt~\cite{synthetixwhitepaper} is a platform
that allows users creating and exchanging synthetic versions of assets such as
gold, silver, cryptocurrencies, and traditional currencies. 
The purpose of \synt is to allow the creation of ``synthetic assets'' whose
prices can track currencies, cryptocurrencies and commodities. 
\synt involves two distinct types of token.
Users first purchase and lock up \synt's native token SNX
into the \synt contract that gets staked
as collateral to back other Synths tokens.
\footnote{The C-Ratio of \synt is 800\% as for now.}
These Synths are synthetic assets created through \synt platform.
Please note that \synt platforms always values sUSD, one of
synthetic assets, at one USD.
The price of Synths is determined through oracles
that report external real-world price of asset to the aggregator
contract which then proceeds the median calculation.
%
As shown in Figure~\ref{fig:synthetix}, the current oracles
and aggregators are provided through \cl.
Each asset type provides an independent \cl's aggregator which
maintains a number of oracle sources.
To ensure accurate data feeds, the oracles update the on-chain price 
in a short period, like 5 or 10 minutes.
%

%% file: fig/valid-range-ample.tex
\begin{figure}[!h]

    \begin{center}
        \begin{tikzpicture}[scale = 0.8,>=stealth]
        
        \draw (0, 0) rectangle node {\scriptsize $\texttt{\ldots}$} +(0.5, 0.5);  

        \draw (0.5, 0) rectangle node {\scriptsize $\texttt{Too Old}$} +(1.5, 0.5);  

        \draw (2, 0) rectangle node {\scriptsize $\texttt{Valid Timestamp}$} +(4, 0.5);  

        \draw (6, 0) rectangle node {\scriptsize $\texttt{Too Recent}$} +(1.8, 0.5);  

        \draw (7.8, 0) rectangle node {\scriptsize $\texttt{\ldots}$} +(0.5, 0.5); 

        \draw[->, thick] (2, -1) -- (2, -0.6);
        \node[scale = 0.8] at (2, -1.2) {$T_0 - expire$};

        \draw[->, thick] (6, -1) -- (6, -0.6);
        \node[scale = 0.8] at (6, -1.2) {$T_0 - delay$}; 

        \draw[->, thick] (8.3, -1) -- (8.3, -0.6);
        \node[scale = 0.8] at (8, -1.2) {$T_0$};
        
        \draw[dotted] (0,0) -- (0, -1);
        \draw[dotted] (2,0) -- (2, -1);
        \draw[dotted] (6,0) -- (6, -1);
        \draw[dotted] (8.3,0) -- (8.3, -1);

        \node[scale = 0.7] at (5.15, -0.25) {\texttt{$expire$}};
        \draw[->] (5.15-0.7, -0.25) -- (2, -0.25);
        \draw[->] (5.15+0.7, -0.25) -- (8.3, -0.25);

        \node[scale = 0.7] at (7.15, -0.5) {\texttt{$delay$}};
        \draw[->] (7.15-0.6, -0.5) -- (6, -0.5);
        \draw[->] (7.15+0.6, -0.5) -- (8.3, -0.5);
        
        \end{tikzpicture}
        \caption{The valid range of a report in \af.}
        \label{fig:validreportrangeample}
    \end{center}
\end{figure}

%% file: sec/measurement.tex
In this section, we present details and results of our measurement studies.
We focus on the 
\af, \synt, \maker, and \cpd platforms, and we measure and report on the following: 
\begin{inparaenum}[a)]
    \item market price volatility of the platforms' assets 
        (Section~\ref{sec:price-volatility}),
	\item deviations between the market prices and prices reported by
        oracles (Section~\ref{sec:price-deviation}),
    \item anomalies that may indicate oracle malfunctions or misbehaving
        (Section~\ref{sec:unstableevents}),
	\item transaction graphs of oracles showing their interactions with the
        ecosystem (Section~\ref{sec:transaction-graph-analysis}).
\end{inparaenum}

\begin{figure*}[bt!]
\centering
\includegraphics[width=0.9\linewidth]{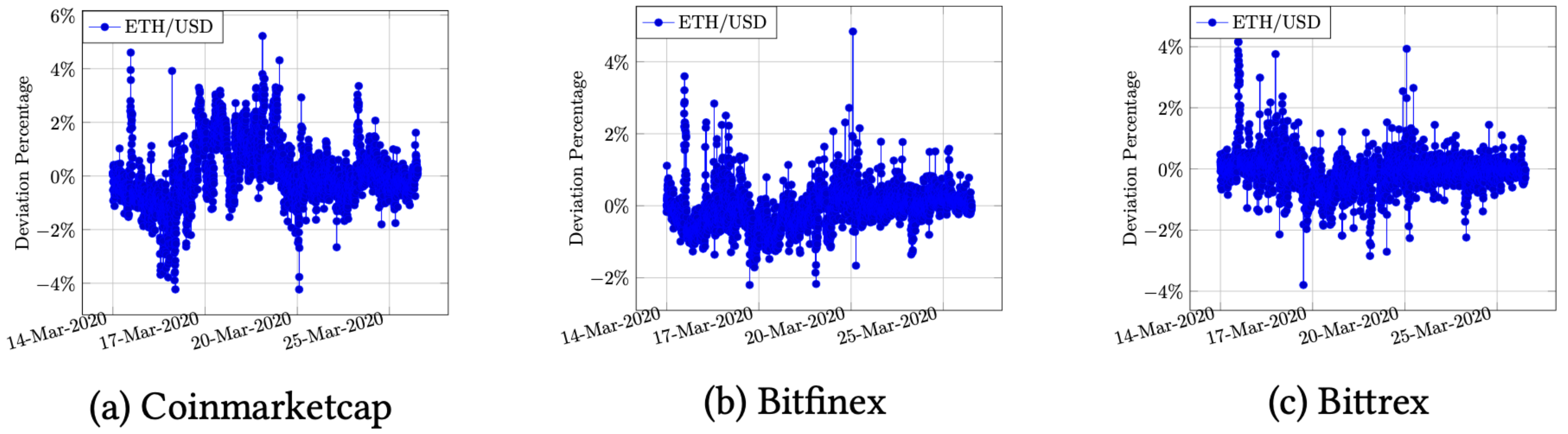}
\caption{Price deviations of the \synt oracle from its price sources.}
\label{fig:deviation-synthetix}
\end{figure*}
\subsection{Price Volatility}
\label{sec:price-volatility}
\begin{table}[h!]
  \caption{Price volatility of DeFi assets.}
  \label{table:dailypricechange}\centering
\footnotesize
\begin{tabular}{lrrrrr}
\toprule
\multirow{2}{*}{Assets} & \multirow{2}{*}{Obs.} & \multicolumn{4}{c}{Daily change}\\ 
\cline{3-6}
 & & $\leq$1\% & $\leq$5\% & $\leq$10\% & >10\%\\\toprule
SAI    &261    &183 (70\%)   &72 (28\%)  &6 (2\%)   &0  \\ 
DAI  &764 &502 (66\%) &253 (33\%) &8 (0.9\%) &1 (0.1\%)\\
AMPL &216 &65 (30\%) &92 (42\%) &32 (15\%) &27 (13\%)\\
SNX & 687 & 78 (11\%) & 272 (39\%) & 213 (31\%) & 124 (19\%)\\
\bottomrule
\end{tabular}
\end{table}
In this section, we demonstrate the price volatility of DeFi assets that aim at removing volatility.
We summarize daily changes of
the market prices (in USD, as reported by \seqsplit{https://coinmarketcap.com}) for each discussed platform in
Table~\ref{table:dailypricechange}, where \texttt{Obs.} is the number of observations (i.e., days that a platform is in operation).
It can be observed that despite of aiming a stability, 
all platforms experienced 1\% or 5\% price changes within one single day.
Moreover, in around 30\% of the investigated days, the market price of AMPL
has more than 10\% daily price change. 

\begin{figure}[t!]
\centering
  \includegraphics[width=1.0\linewidth]{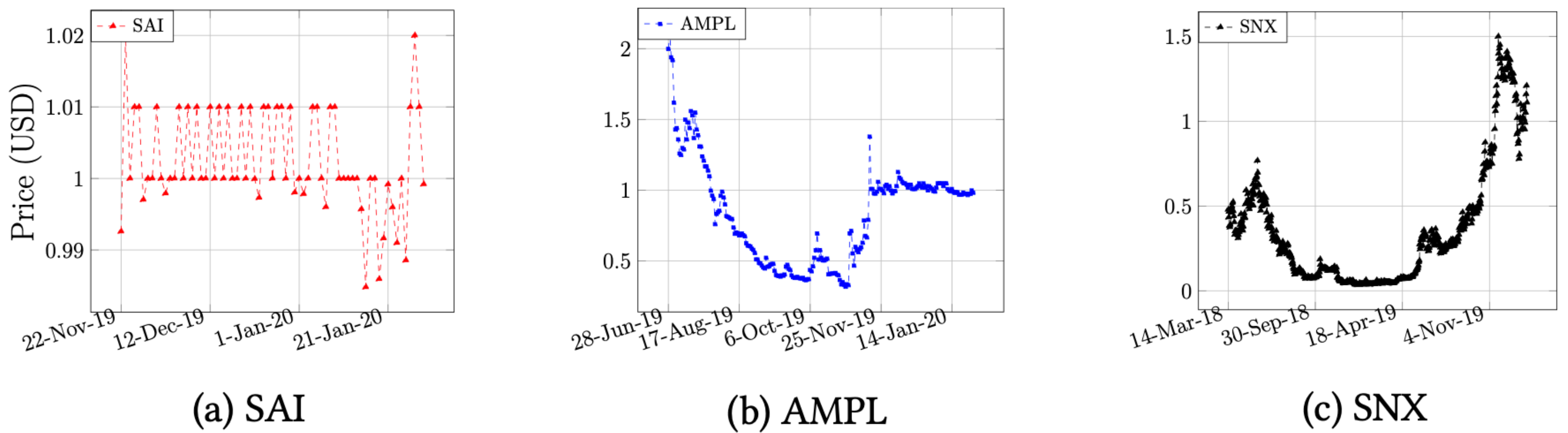}
    \caption{Historical prices of selected DeFi assets.}
\label{fig:market-price}    
\end{figure}
The price volatility over time is depicted in Figure~\ref{fig:market-price}, where
rapid prices changes are mainly caused by changing volume, external events
(like banning cryptocurrencies by countries), or speculation.  
All the results
presented indicate that these DeFi protocols and protocols relying on their
assets require real-time accurate reference price data to hedge the risk of
high volatility.
Therefore, the price oracles reporting accurate price play a crucial role
in the DeFi ecosystem.

\subsection{Price Deviation}
\label{sec:price-deviation}
In this section, we measure the deviation between real-time market prices and
prices reported by the oracles of four major DeFi platforms.  We also investigate
the possible reasons for ``outliers'' -- oracle reports with unusually higher
deviation than other reports.
To conduct the study, we select active oracles, with most frequent reporting, from \maker, \cpd, and \synt, reporting ETH/USD rates.
For \af,  
we investigate its official market oracle\footnote{https://www.ampleforth.org/dashboard/oracles}
which is supposed to report an AMPL/USD rate every 12 hours.
We use  Ethereum's BigQuery database~\cite{ethbigquery} to get data about oracle interactions with their DeFi platforms. 
For each oracle, we analyze its all transactions
by extracting their data and metadata, parsing the data to a readable price format, and
comparing it with the real-world price sources that the oracle is supposed to be following (oracles may use different price sources).
In our experiments, we consider an oracle's claimed source as its baseline for the given asset price.
Moreover, as the methodologies of price reporting by oracle are not strictly specified,
for each baseline source, we also show its real-time ``raw'' price and median values over 1, 5, 10, and 60 minutes.

\begin{figure}[t!]
\centering
\includegraphics[width=1.0\linewidth]{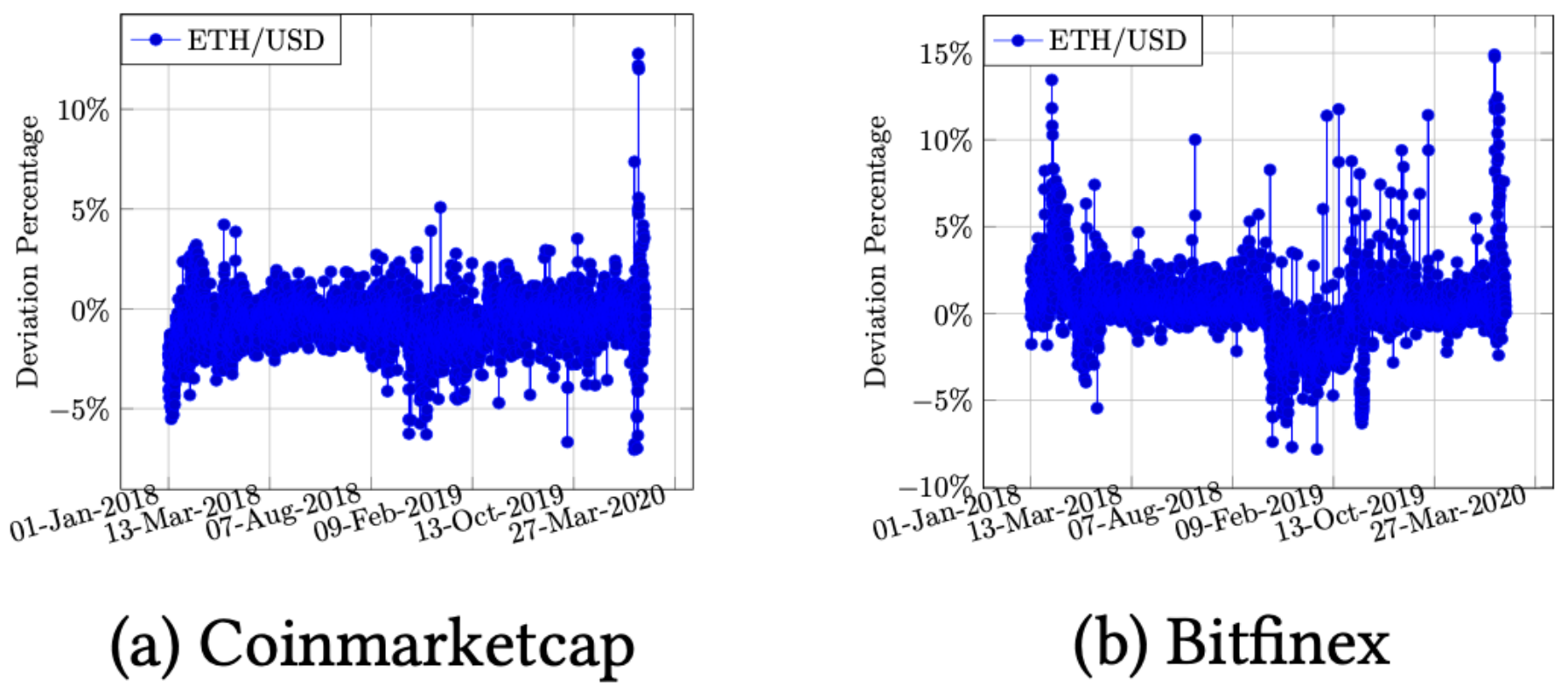}
\caption{Price deviations of the \maker oracle from its price sources.}
\label{fig:deviation-makerdao}
\end{figure}
\point{Results}
We first investigate the \synt oracle,\footnote{Address: 0xac1ed4fabbd5204e02950d68b6fc8c446ac95362} by analyzing its 3,308 price posting and comparing them with the data from different exchanges.
As \synt is integrated with \cl, 
we find that the oracle's claimed ETH price sources~\cite{chainlinkethsource} 
are Coinmarketcap, Bitfinex~\cite{bitfinex} and Bittrex~\cite{bittrex}.
Therefore, in Figure~\ref{fig:deviation-synthetix} we show the ETH/USD price
deviations between the oracle reports and its price sources.
As we can see, the number of deviations is substantial with most deviations
standing within $\pm$2\% range.

Similarly, in Figure~\ref{fig:deviation-makerdao}, we illustrate deviations of the \maker ETH/USD oracle\footnote{Address: 0xfbaf3a7eb4ec2962bd1847687e56aaee855f5d00}.
As the oracle does not specify its sources,
we use the same baselines as in \synt for evaluation, except for Bittrex which
provides the information of ETH/USD rate only since June
2018~\cite{ethbittrex} (we measure the oracle interactions starting from Jan
2018).
As we can see, there is a substantial number of deviations, with most of them
being within the 5\% range, indicating that the \maker oracle is less effective
than the previous \synt oracle.
Moreover, there are a few outlier reports, deviating more than 10\% (we
investigate them further in Table~\ref{table:outliers-analysis}).

\begin{figure}[h!]
\centering
\includegraphics[width=1.0\linewidth]{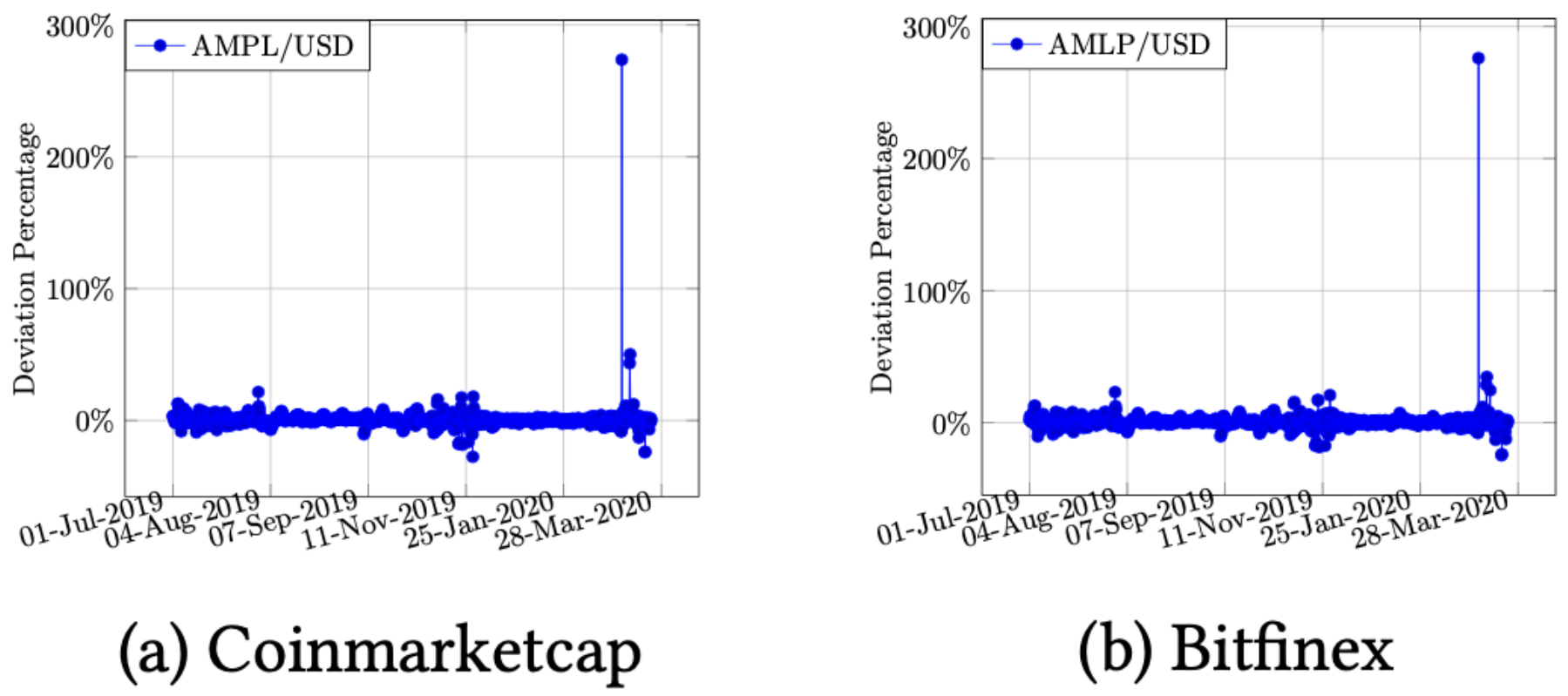}
\caption{Price deviations of the \af oracle from its price sources.}
\label{fig:deviation-ampleforth}
\end{figure}
In \af, the declared sources of oracle
is Anylockanalytics~\cite{anyblock,afsource},
however, it 
does not provide a public API to retrieve real-time prices for
individuals.
Therefore, we consider the same baselines as in \synt, except
for Bittrex which does not track AMPL/USD rates.
We analyze 980 transactions for 
the \af oracle\footnote{Address: 0x8844dfd05ac492d1924ad27ddd5e690b8e72d694}
and check the results against Coinmarketcap and Bitfinex.
As shown in Figure~\ref{fig:deviation-ampleforth}, the majority of deviations are 
within the 5\% range, similarly as for the \synt oracle.
Interestingly, there is a single outlier report with an extremely large deviation value
(i.e., 273.7\%), which we discuss further in this section.

\begin{figure}[h!]
\centering
\includegraphics[width=1.0\linewidth]{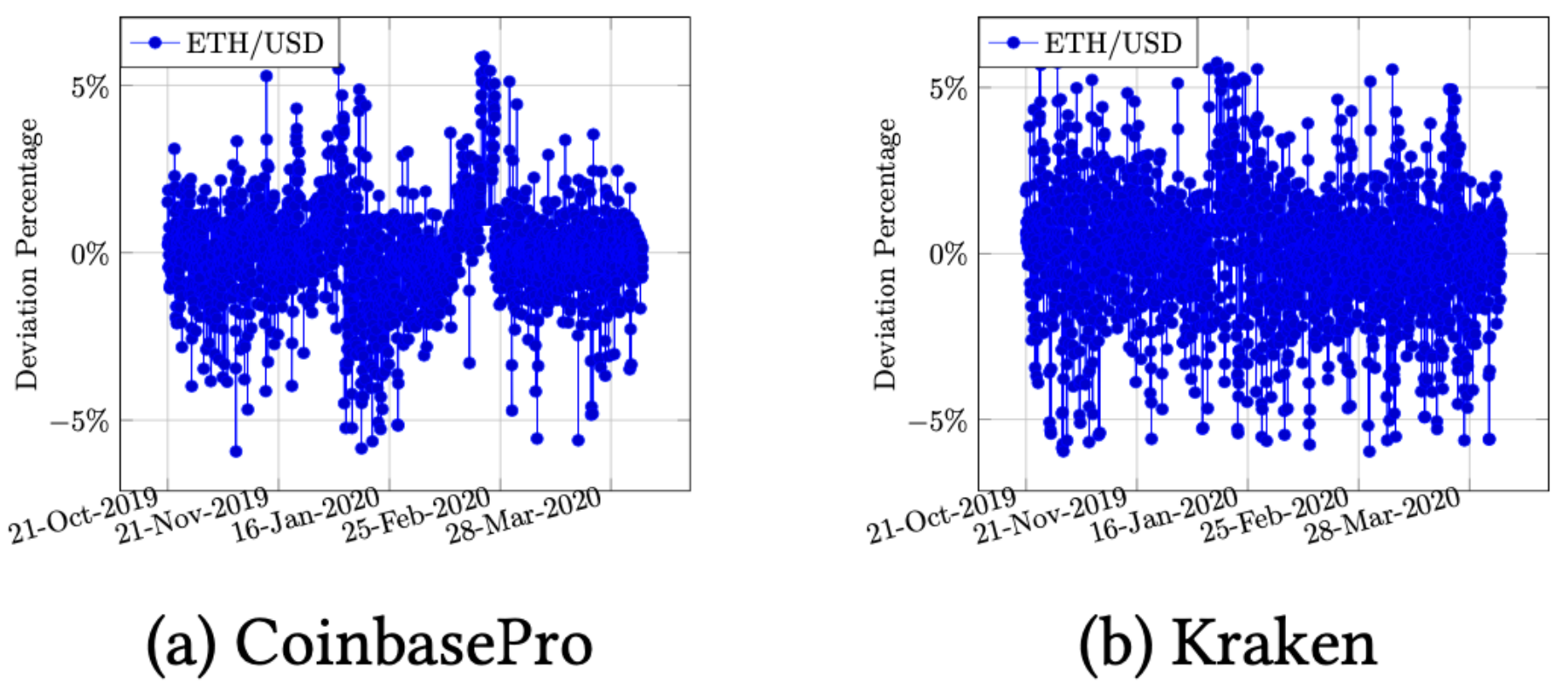}
\caption{Price deviations of the \cpd oracle from its price sources.}
\label{fig:deviation-compound}
\end{figure}

The oracle in \cpd use Kraken~\cite{kraken}
and CoinbasePro~\cite{coinbasepro} as its ETH/USD sources~\cite{compoundclaimed}.
We conduct the evaluation of the \cpd oracle\footnote{Address:
0x3c6809319201b978d821190ba03fa19a3523bd96}, analyzing its 
2,144 transactions in total, 
and presenting the obtained results in Figure~\ref{fig:deviation-compound}.
As we can see, most of the deviations are below the range of 4\%, and there are
only a few deviations above 5\%.

\input{tab/deviation-comparison}
\input{tab/outliers-analysis}
\begin{figure*}[bt!]
\centering
  \includegraphics[width=0.9\linewidth]{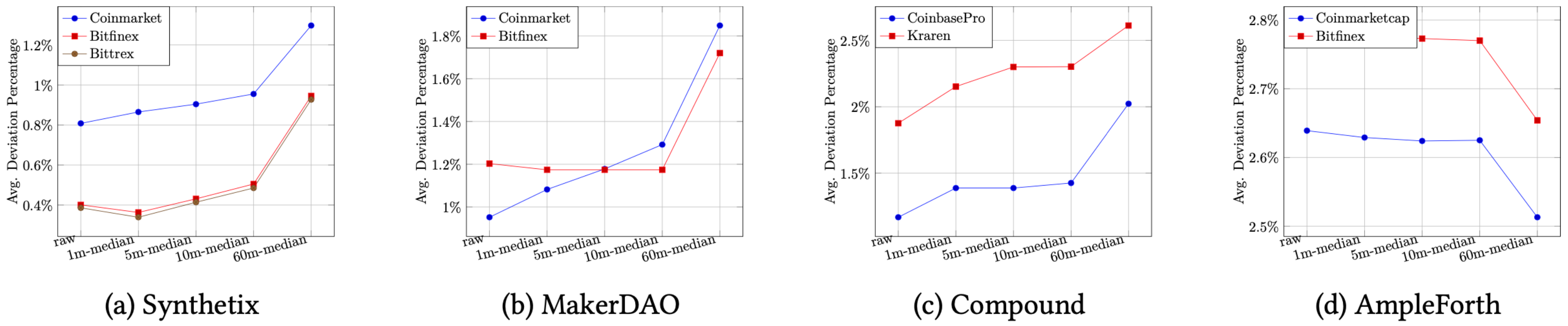}
\caption{Deviation comparison (per platform) among different sources.} 
\label{fig:multiple-deviation}
\end{figure*}
\point{Comparison}
In Figure~\ref{fig:multiple-deviation}, we show the average deviation of each
source for all platforms. 
Please note, that the average deviation is calculated as
$D = \sum_{n=1}^{N} (|D_n|) / N$, where $D_n$ is the percentage of each point
(i.e., transaction)
and $N$ is the sum of points (transactions).
In most cases, the average deviation is below 2\%, which given the volatility of
cryptoassets, can be seen as
relatively precise information.
From Figure~\ref{fig:multiple-deviation}(a) to Figure~\ref{fig:multiple-deviation}(c),
we can see the average deviation increasing
from raw data to 60-minute median. 
In \synt, Bittrex is a more precise source compared to Coinmarketcap and
Bitfinex.
In \maker, Bitfinex is more accurate for the real-time price, 
1-minute, and 5-minute median,
while Coinmarketcap is better for the 10- and 60-minute median.
For \cpd, it is observed that CoinbasePro is much more accurate than Kraken.
Furthermore, the results of the \af oracle indicate an opposite trend to the other three platforms.
That is most likely caused by this oracle processing the price averaged over a longer period
of time before reporting to an aggregator.

To illustrate the differences between the oracles better, we also give the specific numbers in
Table~\ref{table:violateband}. 
We can observe that the deviations of most posting behaviors are $\leq$1\%
and $\leq$5\% except for \af, where its oracle introduces relatively
higher deviation.
A possible reason may be that the baselines measured by us are different from
used by them (as mentioned before, \af oracles do not reveal their price
sources).
However, the average deviation in our measurement is around 2.5\%, which
compared with other platforms seems to be tolerable.

\point{Outliers}
As every oracle may face some inevitable outliers due to
sudden changes in real-time price or mistakes by oracles itself, 
in 
Table~\ref{table:outliers-analysis}
we present selected reports with large deviation values observed in
\autoref{sec:price-volatility}.
In \af, the market oracle had one evident posting error on 5 March
2020\footnote{Tx info: https://bit.ly/2KHiTFE},
when the oracle submitted the price with the hex format 
of \texttt{0x5667f2bb31e073c7} which introduced 273.7\% deviation from the
current exchange price. 
We have not found any reason for this anomaly and we suspect an input
error.\footnote{When changing the first digital number of the
transaction payload, the deviation lowers to 2.9\%, which is a standard range
for this oracle.}
Another interesting anomaly report had 50.2\%
deviation.\footnote{Tx info: https://bit.ly/2K5kSDF},
This inconsistent input was most likely due to the sudden drop in the exchange rate,
hitting the lowest price in the past four months.
The similar situation happened to the oracle of \maker submitting two reports,
deviating 12.8\% and 12.2\% respectively, due 
to the sudden drop of cryptoasset exchange rates in the past three months\footnote{Tx info: https://bit.ly/3ep74BO, https://bit.ly/2K3NcGb}.
In \synt and \cpd, the percentage of top two outlier reports is much smaller
than for the previous two platforms, with just around 5\%.
The largest outlier in \synt
also comes from the sudden drops of real price\footnote{Tx info: https://bit.ly/34Av0xo} while the second largest
is most likely due to the recent US stock market volatility.
The operators of DeFi platforms do not expect large outliers by oracles, therefore 
a well-design accountability mechanism should be introduced, incentivizing 
oracles to report price with promised and tolerable deviation. 
We propose a demonstrative example and will discuss it in
Section~\ref{sec:recommendations}.

\subsection{Failures}
\label{sec:unstableevents}
In this section, we investigate oracle failures. For \maker, \cpd and \af, we
check all transactions which were submitted by their oracles but which
got processed by the Ethereum network unsuccessfully (either rejected by the
network or reverted by the oracle itself).
For \synt, due to the integration with \cl, 
we inspect the oracle nodes of all supported assets, find out the real sources 
they collect from, and then measure  those oracles.
\point{\maker}
The reference price ETH/USD in \maker is updated by an aggregator, 
which collates price data from a number of external sources.
\begin{table}[!t]
  \caption{Oracle failures.}
  \label{table:unstable}\centering
    \footnotesize
\begin{tabular}{lllrrc}
\toprule
Platform                  & Address     &Type       &Total &Failures   & Remark \\
& &(/USD)       &&& \\ \toprule
    \multirow{4}{*}{\maker} &0x000d..&ETH&7042&54 (0.77\%)&out of gas \\
                        &0x005b..&ETH&7545&164 (2.17\%)&out of gas  \\
                        &0x0032..&ETH&11032&154 (1.39\%)&out of gas  \\ 
                        &0xa8eb..&ETH&12633&67 (0.53\%)&out of gas  \\\midrule
    \multirow{2}{*}{\af} &0xd035.. &AMPL&953&30 (3.15\%)&reverted  \\
                     &0xcaef.. &AMPL&295&34 (11.53\%)&reverted  \\
\bottomrule
\end{tabular}
\end{table}
As shown in Table~\ref{table:unstable}, 
there are 54 failed transactions occurred in the total 7,042 due to 
\textit{out-of-gas} error\footnote{This error happens when a transaction
to be completed requires more computing resources than provided by its sender.} 
in one ETH oracle\footnote{Address: 0x000df128eb60a88913f6135a2b83143c452c494e}, 
with 0.77\% failure rate. 
The similar issues can be found for the another three
oracles\footnote{Addresses: 0x005b903dadfd96229cba5eb0e5aa75c578e8f968,
0x0032ad8fae086f87ff54699954650354bb51e050, 0xa8eb82456ed9bae55841529888cde9152468635a}
with 2.17\%, 1.39\% and 0.53\% failure rates, respectively.

\point{\af} 
Next, we investigate all transactions 
originated from \af's oracles.
In Table~\ref{table:unstable}, 
we show our findings that 30\footnote{Address: 0xd0352aad6763f12d0a529d9590ea2f30421667a6} and 
34\footnote{Address: 0xcaefaf2130f0751520d5a6a62f3b9c9eaa4739f4} reverted transactions happened in two market oracles 
of \af by April 2020.

\point{\synt} 
\synt is fully integrated with \cl for feed services.
Each asset type supported in \synt has a collection of corresponding \cl's
nodes to complete oracle-like jobs.
We further review the nodes for all exchange pairs -- 
ETH/USD, BTC/USD, AUD/USD, EUR/USD, CHF/USD, GBP/USD, JPY/USD, XAG/USD and XUG/USD,
to find out potential failures.

\input{tab/synthetixoracle}
In Table~\ref{table:synthetixandchainlink}, we give the details on oracles, their
respective sources, and the issues encountered.
We found that Omniscience, Ztake.org, Anyblock, and Simply VC submitted
transactions which subsequently have been reverted by the Ethereum network.
By 14 February 2020, Alpha Vantage received 17 requests from
\cl but ignored them, submitting no successful responses.
LinkPool employs two external sources, i.e., CryptoCompare and Alpha Vantage,
for ETH, BTC, AUD, and XAG rates. However, Alpha Vantage
is unreliable which renders LinkPool to be unreliable too.
Fiews, Cosmostation, Validation, etc,. are stable nodes with no anomalies found,
while the sources of stake.fish and Chainlayer are unknown to the public, thus
cannot be publicly audited.

\subsection{Transaction Activity Analysis}
\label{sec:transaction-graph-analysis}
An oracle address may interact with a large number of Ethereum addresses
that could be an ERC-20 token contract, an on-chain service, an entity
from other protocol or an external account address, etc.
We should suspect the trustworthiness of those addresses that
oracle interact with 
since oracles should be trustworthy enough to achieve high accountability.
In this section, we focus on the transaction activities analysis for oracles
of the DeFi platforms.
We crawl the entire transaction history of oracles by using BigQuery, 
then we build transaction graphs, and find what are most common addresses
oracles interact with, 
what entities or external accounts
they communicate with, and what interesting activities they are involved in.

\point{\af}
We collect 132,119 transactions from the market oracle of \af, and  
find out that there are 47 different addresses interacting with the oracle.
The large proportion of entire transactions are interactions with \cl's aggregators.
The market oracle has 161 transactions with
\texttt{UpgradeProxy} contract of \af to set or update certain parameters.
Others are external account addresses with 48 and 1 transaction
involved. The oracle sent 48 transaction without input data to
an external account\footnote{Address: 0x43eb83a6b54a98b2d051c933b8e4a900d6bacbee}
in succession on 13 March 2020 (most likely, these are testing activities).

\point{\maker}
Similarly, we use the ETH/USD \maker oracle as 
our measurement target,
extract 4,914 transactions, and analyze them.
The oracle interacts with four types of entities from
seven different addresses.
Most transactions are about price posting behaviors, however,
there are two failed one due to
out-of-gas error.
It has four proxy activities and only three token transfers
of SAI and DAI.
Moreover, there is one migration activity when the platform
decided to make SAI and DAI conversion.

\point{\cpd}
We select ETH/USD price oracle in \cpd and analyze its all 11,458 transactions.
All transactions are about the price reporting actions interacted with three 
on-chain aggregators in total.
In contrast to the oracles of other platforms, the transaction history
does not contain interactions with other actors or services.

\point{\synt}
Similar to other platforms, most of the 142,422 transactions 
of the \synt ETH/USD oracle are interactions with the active 
\cl's aggregator contracts.
An exception is one of the aggregators\footnote{Address:
0x5c545ca7f9d34857664fdce6adc22edcf1d5061f} which is self-destructed
without a clear reason.
Besides that, the oracle is also involved in 2,056 transactions with the \synt network
contract, most are about retrieving the value of parameters from the platform.
One interesting activity is 667
transactions sent in total to itself without input data (most likely for testing
purposes).

In conclusion, only the price oracle in \cpd is not involved in any other
activities except reporting reference price. 
We encourage the operator of price oracle to maintain high accountability
by keeping the least interactions with external services or actors.

%% file: tab/deviation-comparison.tex
\begin{table}[!h]
  \caption{Deviation comparison between different platforms.}
  \label{table:violateband}\centering
  \footnotesize
\begin{threeparttable}
\begin{tabular}{L{1.25cm}C{0.5cm}R{0.7cm}R{0.7cm}R{0.5cm}R{0.5cm}R{0.35cm}R{0.35cm}R{0.6cm}}
\toprule
\multirow{2}{*}{Oracle}&\multirow{2}{*}{Obs.}&\multirow{2}{*}{Source}&\multirow{2}{*}{Rule}&\multicolumn{4}{c}{\# of dev. in range (\%)}&\multirow{2}{*}{\shortstack[l]{Avg d-\\ev (\%)}}
\\\cline{5-8}
& &&&$\leq$1&$\leq$5&$\leq$10&>10&\\\toprule
\multirow{12}{*}{\shortstack[l]{Synthextic \\(ETH/USD)}}&\multirow{12}{*}{3308}&\multirow{4}{*}{CMC\tnote{1}}&raw&2388&969&1&0&0.808\\
& & &1m&2245&1061&2&0&0.865\\
& & &10m&2115&1183&10&0&0.955\\
& & &1h&1733&1502&72&1&1.297\\\cline{3-9}
& & 
\multirow{4}{*}{BF\tnote{2}}&raw&3088&220&0&0&0.400\\
& & &1m&3159&149&0&0&0.363\\
& & &10m&2914&393&1&0&0.505\\
& & &1h&2227&1046&35&0&0.945\\\cline{3-9}
& & 
\multirow{4}{*}{Bittrex}&raw&3048&260&0&0&0.386\\
& & &1m&3121&186&1&0&0.339\\
& & &10m&2925&381&2&0&0.485\\
& & &1h&2276&997&35&0&0.927\\\cline{1-9}
\multirow{8}{*}{\shortstack[l]{MakerDAO \\(ETH/USD)}}&\multirow{8}{*}{4707}&\multirow{4}{*}{CMC}&raw&2974&1708&22&3&0.952\\
& & &1m&2733&1928&43&3&1.082\\
& & &10m&2404&2226&73&4&1.292\\
& & &1h&1799&2649&231&28&1.849\\\cline{3-9}
& & 
\multirow{4}{*}{BF}&raw&2903&1679&108&17&1.203\\
& & &1m&2813&1791&93&10&1.174\\
& & &10m&2813&1791&93&10&1.174\\
& & &1h&1991&2472&224&20&1.720\\\cline{1-9}
\multirow{8}{*}{\shortstack[l]{Compound \\(ETH/USD)}}&\multirow{8}{*}{2144}&\multirow{4}{*}{CBP\tnote{3}}&raw&1309&816&19&0&1.168\\
& & &1m&1207&859&78&0&1.388\\
& & &10m&1121&933&89&1&1.426\\
& & &1h&871&1084&153&36&2.023\\\cline{3-9}
& & 
\multirow{4}{*}{Kraken}&raw&1059&1030&55&0&1.876\\
& & &1m&1020&1055&69&0&2.152\\
& & &10m&1001&992&151&0&2.302\\
& & &1h&979&989&143&33&2.612\\\cline{1-9}
\multirow{8}{*}{\shortstack[l]{AmpleForth \\(AMPL/USD)}}&\multirow{8}{*}{980}&\multirow{4}{*}{CMC}&raw&418&449&86&27&2.639\\
& & &1m&411&458&84&27&2.629\\
& & &10m&413&451&83&28&2.625\\
& & &1h&448&423&83&26&2.513\\\cline{3-9}
& &
\multirow{4}{*}{BF}&raw&387&455&112&26&2.792\\
& & &1m&388&455&110&27&2.777\\
& & &10m&397&444&112&27&2.770\\
& & &1h&417&430&107&26&2.654\\
\bottomrule                                          
\end{tabular}
\begin{tablenotes}[para,flushleft]\footnotesize
    \item[1] Coinmarketcap.
    \item[2] Bitfinex.
    \item[3] CoinbasePro.
  \end{tablenotes}
\end{threeparttable}
\end{table}

%% file: tab/outliers-analysis.tex
\begin{table*}[!tb]
  \caption{Outliers analysis.}
  \label{table:outliers-analysis}\centering
    \footnotesize
\begin{tabular}{L{1.3cm}R{1.2cm}R{2.4cm}R{1.5cm}R{1.7cm}R{1.3cm}C{6.2cm}}
\toprule
Platform & Tx hash & Time &Oracle (/USD) & Baseline (/USD) & Dev. (\%) &
    Possible explanations \\\toprule
\multirow{2}{*}{AmpleForth}
    &0x9c61..&2020-03-05 00:31&6.23&1.66&273.7&Oracle wrong input (most likely)\\
&0x67fb..&2020-03-13 00:39&0.855&0.569&50.2&Exchange price sudden drop\\\midrule
\multirow{2}{*}{Compound}
&0x33ba..&2020-02-17 16:02&271&256&5.85&Exchange price sudden drop\\
&0xde2f..&2019-11-08 18:12&181&191&-5.9&Exchange price sudden increase\\\midrule
\multirow{2}{*}{MakerDAO}  &0x698a..&2020-03-13
    02:44&121.49&107.73&12.8&Exchange price sudden drop\\
                           &0x38b0..&2020-03-13 02:36&110.43&98.43&12.2&Exchange
                           price sudden drop\\\midrule
\multirow{2}{*}{Synthetix}&0xd9f0.. &2020-03-19 16:29&138.66&131.78&5.2&Exchange
    price sudden drop\\
                           &0x3860..&2020-03-15 00:30&127.82&122.19&4.6&US stock
                           market volatility\\
\bottomrule    
\end{tabular}
\end{table*}

%% file: tab/synthetixoracle.tex
\begin{table*}[!h]

  \caption{The price oracles of Synthetix.}
  \label{table:synthetixandchainlink}\centering
  \footnotesize
\begin{tabular}{L{2.4cm}L{4.5cm}C{4.1cm}}
\toprule
    Oracle                 & Price (/USD) & Issues \\ \toprule
Omniscience            & ETH, GBP, XAU               & Reverted transaction(s)\\ \midrule
Anyblock               & BTC                  & Reverted transaction(s)\\ \midrule
Ztake.org              & AUD, BTC, XAG & Reverted transaction(s)\\ \midrule
Simply VC              & BTC, EUR, CHF, GBP, XAG & Reverted transaction(s)\\ \midrule
Alpha          & AUD, EUR, CHF, GBP, JPY & 17 requests
    from ChainLink but 0 successfull responses \\ \midrule
LinkPool               & ETH, BTC, AUD, XAG & Two sources: CryptoCompare and Alpha Vantage.                \\ \midrule
stake.fish & ETH, AUD, JPY & Oracles are unknown to the public\\\midrule
Chainlayer & ETH, AUD, JPY & Oracles are unknown to the public\\\midrule
Fiews                  & EUR, CHF, JPY, XAG  & N.A.\\ \midrule
Cosmostation           & ETH  &N.A. \\ \midrule
Validation             & ETH, CHF  &N.A. \\ \midrule
SDL       & BTC, EUR   &N.A. \\ \midrule
LinkFprest             & CHF  &N.A. \\ \bottomrule
\end{tabular}
\end{table*}

%% file: sec/discussion.tex
\subsection{Decentralization}
\label{sec:decentralization}
In the background section, we discuss different designs of DeFi oracles.
Some of them rely on centralized aggregators to retrieve reference
price while others develop the partnership with \cl's feed providers.
In this section, we investigate how oracles systems are implemented in practice
and how that can influence the aimed decentralization of the platforms.

\begin{table}[!tb]
  \caption{Comparison of the oracle designs.}
  \label{table:decentralized}\centering
  \footnotesize
\begin{tabular}{lrllrr}
\toprule
Platform  & Freq. & Design & Hierarchy \\
& (/hour) & & \\ \toprule
\maker    &1, 6         &Centralized aggregator                     &Centralized      \\ 
\cpd      &2, 5            &Centralized aggregator                    &Centralized  \\
\af       &12               &Aggregator + \cl    &Semi-decentralized   \\
\synt     &1/12             & \cl  nodes       &Decentralized  \\
\bottomrule
\end{tabular}
\end{table}
Table~\ref{table:decentralized} describes the selected properties of oracles that
influence their decentralization.
\maker and \cpd have similar architectures -- they employ one single aggregator 
to periodically retrieve price information from external whitelisted oracle nodes.
This design introduces inherent centralization drawbacks even though other
components of these systems
are deployed on decentralized smart contract platforms.
\af employs \cl to provide oracle functionalities, which also mitigates (due to
the design of \cl) the centralized risks of a single aggregator.
However, it still relies on an aggregator contract to collect data from four oracles.
In a near future, \af is planned to be fully integrating with \cl for data feeds,
thus, we classify it as a semi-decentralized design as of now.
\synt has announced that \synt-\cl integration is now operational on Ethereum~\cite{clwithsynt}, providing fully decentralized price feeds. 
The data feed will be offloaded to the decentralized oracle network of \cl and
the reference price rates are transferred on-chain by a number of 
independent nodes backed by economic incentives rather than any central parties.
Therefore, as for now, its design is the closest to being decentralized.

\subsection{Recommendations}
\label{sec:recommendations}
Our study, which can be seen as initial, indicates that the oracle ecosystem is
immature. Therefore, in this section, we try to learn from our observations and
give insights on the potential improvements of future oracle platforms.

\point{Transparency}
As discussed in Section~\ref{sec:price-deviation}, the methodologies of price
processing by oracles are not clearly stated.  Even the sources that the oracle
retrieves from are ambiguous or unknown to the platform users.  This results in
a lack of transparency and potentially undetectable misbehavior of oracle
platforms as currently no entity is able to provably examine the precision of
price reported by oracles.  
Our first recommendation for future oracle designs
is to require oracles to explicitly declare their manifest,
and we give a demonstrative example in 
Algorithm~\ref{alg:manifestcontract}.
Such a manifest
would contain oracle metadata (like oracle contact information), deployed data
sources, intended frequency of oracle updates, and precise description of the
price derivation.  Due to its properties, we see the underlying blockchain
platform as a natural place of publishing such manifests.
First, the operator of an oracle deploys a manifest contract. The contract
describes the main properties of the oracle: its owner, data sources,
the publishing frequency, and the data precision.
The manifest makes the oracle operations more transparent, since the relying
parties have better guarantees about the data feeds and can audit the oracle
in an easier way.
We also emphasize, that the manifest does not make the oracle itself more
vulnerable.
\begin{algorithm}[!t]
    \caption{Manifest Contract.}
    \label{alg:manifestcontract}
    \SetAlgoVlined
    $address~owner,~string[]~sources$\\
    $unit256~frequency,~float~precision$\\
    \medskip
    \texttt{/*~{\it sources}: the claimed data sources~*/}\\
	\texttt{/*~{\it frequency}: intended frequency of updates~*/}\\
	\texttt{/*~{\it precision}: maximum tolerated deviation~*/}\\
    \SetKwProg{func}{function}{}{}
    \func{constructor(src, freq, prec)}{
        owner $\leftarrow$ msg.sender, sources $\leftarrow$ src\\
        frequency $\leftarrow$ freq, precision $\leftarrow$ prec\\
    }

    \func{set(src, freq, prec)}{
        assert(owner == msg.sender), sources $\leftarrow$ src\\ 
        frequency $\leftarrow$ freq, precision $\leftarrow$ prec\\ 
    }
    
    \func{get()}{
        \Return (sources, frequency, precision)
    }
\end{algorithm}

\point{Accountability}
We believe that oracles, becoming critical trusted parties, should be held
accountable for their actions. In the blockchain ecosystem, we can envision that
a feasible way of implementing accountability is crypto-incentives.  Therefore,
to incentivize oracles to report accurate prices in promised frequency, we can
imagine that the platforms implement mechanisms that would punish an oracle
violating its manifest or platform policies, e.g., events like late or missing
reports, or provable misbehavior like a high price deviation.  
%
Manifest contracts, presented in the previous section, allow not only to keep oracles
auditable but also accountable for their actions.
Since with a published manifest some oracle misbehavior can be proved, we
propose an accountability mechanism to punish oracles for such misbehavior.
More specifically, the platform is initiated with an accountability contract
with which oracles deposit some specific amount of cryptoassets.
Whenever an oracle misbehaves, violating its manifest, it can be
punished by the accountability contract which verifies the reported
misbehavior and executes all (or the part of) the deposit.  The details of an
example accountability contract are presented in
Algorithm~\ref{alg:accountablecontract},
and we believe that such a mechanism may be helpful to incentivize oracles to provide
better services.
\begin{algorithm}[!t]
    \caption{Accountability Contract.}
    \label{alg:accountablecontract}
    \SetAlgoVlined
    $mapping(address \Rightarrow \textlangle manifest,~deposit\textrangle)~Oracles$\\
    $struct~report\{unit~timestamp,~float~deviation\}$\\
    \medskip
    \SetKwProg{func}{function}{}{}
    \func{deposit(manifestAddr)}{
        Oracles[msg.sender].manifest $\leftarrow$ manifestAddr;\\
        Oracles[msg.sender].deposit $\mathrel{{+}{=}}$ msg.value;
    }
    
    \func{punish(oracle, report)}{
        \If{!checkMisbehavior(oracle, report)}
        {	
        	msg.sender.transfer(Oracles[oracle].deposit) \\
        	Oracles[oracle].deposit $\leftarrow$ 0
        }
    }
    
    \func{checkMisbehavior(oracle, report)}{
    	\texttt{/*~check late or missing reports~*/}\\
        \If{report.timestamp is late or report is NULL}
        {\Return~~{\it False}}
    	\texttt{/*~check high deviation~*/}\\
		\If{report.deviation $\geq$ Oracles[oracle].manifest.get().precision}
        {\Return~~{\it False}}
        \Return~~{\it True}
    }
    
\end{algorithm}

\input{tab/operational-robustness}
\point{Operational Robustness}
We found it surprising that despite relatively simple oracle interactions, they
are not free from basic operational issues (e.g., causing out-of-gas errors).
Since oracle reports play a crucial role in the DeFi ecosystem, we encourage
operators to overprovision them by increasing their gas and gas price.  The
former has to guarantee that there is enough gas to be consumed by the entire
execution of the transaction.  The latter parameter is important for the latency
of reports, which may be especially important when the Ethereum network is
congested~\cite{gasincreased}.  A high gas price would allow oracle reports to
be appended to the blockchain much faster, since the blockchain network
prioritizes more expensive transactions to be added first.
We empirically study the historical transactions of oracles discussed in 
Section~\ref{sec:price-deviation}, analyze the 
gas cost, gas price and transaction fees required,
and 
%
give insights into how much gas and what gas price should be prepared 
in Table~\ref{table:operational-robustness}. 
As shown, the operator of oracle in \synt requires to prepare around 642,963 on average to 
execute the price reporting transactions.
In \maker, the average gas price is 7.9 Gwei\footnote{Gwei is a unit of ether, 
1 Ether = $10^9$ Gwei}, which is relatively cheap compared with \synt and \af.
In addition, \cpd has the cheapest transaction fees paid by its oracle operators to report price.

%% file: tab/operational-robustness.tex
\begin{table*}[!h]
\caption{Gas and gas price of reporting transactions by oracles in all platforms.}
  \label{table:operational-robustness}\centering
\begin{tabular}{l|r|rrr|rrr|rrr}
\toprule
\multirow{2}{*}{Platform} & \multirow{2}{*}{Obs.} & \multicolumn{3}{c}{Gas} & \multicolumn{3}{c}{Gas Price (Gwei)} & \multicolumn{3}{c}{Tx Fees (USD)} \\\cline{3-11}
& &Min. & Avg. & Max.  &Min. & Avg. & Max. &Min. & Avg. & Max.                 \\\toprule
Synthetix &3308 &585,769&642,963&895,123  &12.657&19.277&150.000  &0.116&1.559&12.084\\\midrule
MakerDAO  &4707 &77,362&138,512&592,276  &5.011&7.928&208.725   &0.005&0.516&13.561\\\midrule
Compound &2144 &34,807&54,736&205,134  &1.200&8.133&270.000    &0.001&0.001&0.002\\\midrule
AmpleForth &980  &134,170&192,365&231,220  &1.300&13.809&150.001  &0.001&0.002&0.017\\
\bottomrule                        
\end{tabular}
\end{table*}

%% file: sec/related.tex
The only directly related study to ours (up to our best knowledge) is presented
by Gu et al.~\cite{guempirical}. They analyze the accuracy of a
\maker price oracle as well as disagreements between reports by oracle peers.
However, unlike our work, their study is limited to the \maker platform, focuses
on older \maker's v1 oracles, and utilizes CryptoCompare~\cite{cryptocompare} as
baseline for the ETH/USD reference price (while we measure the data
sources that are explicitly claimed by the oracles).
Below, we discuss other related topics in the following categories.

\point{Oracle Designs}
Town Crier (TC)~\cite{zhang2016town} is an authenticated data feed system for smart
contracts. The entity TC acts as a bridge between smart contracts and
existing web sites, which are already commonly trusted for
non-blockchain applications. 
It combines a blockchain front
end with trusted hardware (i.e., the Intel SGX technology~\cite{costan2016intel}) back end to scrape HTTPS-enabled websites and serve source-authenticated data to relying upon
smart contracts.
Due to the integration of the SGX enclave,  
it is possible in TC to conduct a remote attestation that the correct code was executed.
TC establishes a secure TLS connection with a website and parses its content, which
then can be used as an input to smart contracts.
However, one potential limitation of TC is it positions Intel
as a trusted party required to execute a remote attestation.

TLS-N~\cite{ritzdorf2017tls} is a general TLS extension that
provides secure non-repudiation to the TLS protocol. 
It modifies the TLS stack such that TLS records sent by
a server are authenticated (in batches). 
Therefore, the clients can present received
TLS-N records to the third parties which can verify it, just trusting the server. 
In general, TLS-N generates non-interactive proofs about the content of a TLS session that can be efficiently verified by third parties
and blockchain-based smart contracts. 
As such, TLS-N increases
the accountability for the content provided on the web and enables
a practical and decentralized blockchain oracle for web content.
However, the main drawback is its deployability. 
It requires significant changes to the TLS protocol and adoption processes are pretty slow. 

Practical Data Feed Service (PDFS)~\cite{guarnizo2019pdfs} is a system that extends content providers by including new features for data transparency and consistency validations. 
It allows content providers to link their web entities with their blockchain
entities.
In PDFS, data is authenticated over blockchain but without breaking TLS trust
chains or modifying TLS stacks. Moreover, content providers can specify data formats
they would like to use freely, thus data can be easily parsed and customized for smart contracts. 
PDFS keeps content providers auditable and mitigates their malicious activities (e.g., data modification or censorship) and allows them to create a new business model.
The shortcoming is the validation logic placed in smart contracts is not 
too lightweight or efficient and the design with shorter proofs could be 
a potential improvement for PDFS.

\point{DeFi Surveys}
\cite{clark2019sok} provides an understandable survey of on
mainstream DeFi protocols, mainly focus on stablecoins designs.
\cite{moin2019classification} and \cite{pernice2019monetary} systematically discuss the
general designs of existing DeFi platforms.
By decomposing the designs
into various components, i.e., peg assets, collateral amount,
price information and governance mechanism, 
such surveys aim to explore the strengths and drawbacks of DeFi platforms to identify future directions.

\point{Attacking DeFi}
\cite{qin2020attacking} presents a detailed exploration of the flash loan mechanism for the DeFi systems. It analyzes two existing attacks vectors with ROIs beyond 500k\%, then formulates finding flash loan-based attack parameters as an optimization problem. 
It also shows how two previously executed attacks can be boosted to result in a profit of 829.5k USD and 1.1M USD, which is a boost of 2.37× and 1.73×, respectively.
Lewis et al.~\cite{gudgeon2020decentralized} discuss over-collateralization and governance attack on \maker, explore how the weaknesses of design
could lead to DeFi crisis, and propose a new financial contagion. 

\cite{xu2019anatomy} investigates 412 pump-and-dump activities and discovers patterns in crypto-markets associated with pump-and-dump schemes by building a model that predicts the pump likelihood of all assets listed in a crypto-exchange prior to a pump. 
Josh et al.~\cite{kamps2018moon} examines existing information on pump-and-dump schemes from classical economic literature to cryptocurrencies, and proposes criteria that can be used to define a cryptocurrency pump-and-dump. The patterns can
exhibit anomalous behavior, are utilized to locate points of anomalous trading activity in order to flag potential pump-and-dump activity.


%% file: sec/conclusions.tex
In this paper, we present a comprehensive measurement study of DeFi oracles.  
We first dispel the
mist of oracle designs of mainstream DeFi platforms.  By conducting large-scale
measurements of deployed oracles for four prominent open DeFi platforms --
\maker, \cpd, \af and \synt, we investigate the details of price deviation that
comes from the differences between the price information provided by real-time
price and oracle nodes.  We compare the price deviations of the deployed
platforms, conduct the detailed measurement on the stability, accountability, and
deployment patterns of oracles.  We find that deviations from the claimed
sources as well as operational failures are quite frequent.
Finally, we give the discussion on the potential security vulnerabilities
that such platforms may suffer from and give recommendations that could improve
some of the found drawbacks.